\documentclass[prl,aps,twocolumn,footinbib]{revtex4}
\usepackage{epsfig}
\usepackage{amsfonts}
\usepackage{latexsym}
\usepackage{amsmath}
\usepackage{amssymb}
\usepackage{slashed}
\usepackage{longtable}

\usepackage{graphicx}
\usepackage{hyperref}
\usepackage{pstool}

\numberwithin{equation}{section}

\newcommand{\bea}{\begin{eqnarray}}
\newcommand{\eea}{\end{eqnarray}}
\newcommand{\be}{\begin{equation}}
\newcommand{\ee}{\end{equation}}
\newcommand{\ba}{\begin{align}}
\newcommand{\ea}{\end{align}}

\def\Or[#1]{{\text{O}}\left({#1}\right)}
\def\dotl[#1,#2]{\left\langle #1, #2 \right\rangle}
\def\dotlb[#1,#2]{[ #1, #2 ]}
\def\dotp[#1,#2]{(#1) \cdot (#2)}
\def\aff[#1,#2]{\hat{#1}(#2)}
\def\n4sym{{\cal N}=4 SYM}
\def\>{\rangle}
\def\<{\langle}
\def\weight[#1,#2,#3]{\{(#1),#2,#3\}}
\def\ads[#1]{$\text{AdS}_{#1}$}

  \makeatletter
  \let\over=\@@over \let\overwithdelims=\@@overwithdelims
  \let\atop=\@@atop \let\atopwithdelims=\@@atopwithdelims
  \let\above=\@@above \let\abovewithdelims=\@@abovewithdelims

\usepackage{xcolor}

\begin{document}

\preprint{}

\title{Berry curvature and Hall viscosities in an anisotropic Dirac semi-metal}

\author{Francisco Pe\~{n}a-Benitez}
\author{Kush Saha}
\author{Piotr Sur\'owka} 
\affiliation{Max-Planck-Institut  f\"ur Physik komplexer Systeme, N\"othnitzer Str. 38, 01187 Dresden, Germany}

\begin{abstract}
We investigate parity-odd non-dissipative transport in an anisotropic Dirac semi-metal in two spatial dimensions. The analysis is relevant for interacting electronic systems with merging Dirac points at charge neutrality. For such systems the dispersion relation is relativistic in one direction and non-relativistic in the other. We give a proposal how to calculate the Berry curvature for this system and use it to derive more than one odd viscosities, in contrast to rotationally invariant systems. We observe that in such a model the odd part of stress tensor is parameterised by two independent transport coefficients and one that is identically zero.
\end{abstract}

\maketitle

{\em Introduction.--}Since the discovery of quantum Hall states, the topological response of these systems continues to be one of the emerging fields of research \cite{avron1995zograf,read2009,taylor2011fradkin,read2011rezayi,stone2012,bradlyn2012read}. In particular, there has been a revived interest in understanding the interplay between geometry and quantum Hall states with fractional and integer fillings \cite{Wiegmann2013,Fremling2014,can2014wiegman,parrikar2014robert,Abanov:2014ula, gromov2015fradkin}. A key quantity that encodes the topological response to the geometry deformations is Hall viscosity \cite{avron1995zograf,1997physics12050Avron} (see \cite{Hoyos:2014pba,Klevtsov:2016bos} for a review); a nondissipative part of the viscosity tensor that is odd under time reversal and hence nonvanishing only in systems without time-reversal symmetry.

When rotational symmetry is broken, the odd part of the two-dimensional viscosity tensor can have three non-zero components, in contrast to usual single viscosity in a rotationally symmetric system. Despite extensive studies on both isotropic two-dimensional (2D) electron gas and Dirac materials \cite{avron1995zograf,read2009,taylor2011fradkin,read2011rezayi,stone2012,bradlyn2012read,sharafati2016vignale,Vozmediano2016TDM}, systems without rotational symmetry have received surprisingly little attention. 
This follows either from the scarcity of physically realizable examples or from the difficulty in the explicit calculation of Berry phases. Although some progress has been made in set-ups, in which the anisotropy is introduced via the mass tensor or interaction tensor \cite{gromov2017bradlyn,Haldane2015} in a 2D electron gas, the anisotropic case in 2D Dirac semi-metals has not been explored so far.

The objective of this letter is to fill this gap by studying Hall viscosity tensor in a new class of 2D anisotropic Dirac semi-metals \cite{banerjee2009pickett,depplace2010montambaux,dietl2008montambaux}. Such semi-metals are known to exhibit a special phase, namely critical semi-Dirac phase, which is characterized by electronic bands touching in a discrete set of nodes about which the bands disperse linearly in one direction and quadratically along the orthogonal direction. The low-energy Hamiltonian describing such materials reads
\be
\label{eq:SemiDirac}
\mathcal{H}=\mathbf{d}(\mathbf{p})\cdot \boldsymbol \sigma,
\ee
where $\boldsymbol \sigma$'s are Pauli matrices. $\mathbf{d}(\mathbf{p})=(\frac{p_x^2}{2m_0}-\delta_0, p_y,0)$ with $m_0$ being a mass and $\delta _0$ the gap parameter.  This type of Hamiltonian has been argued to emerge in TiO$_2$/VO$_2$ heterostructures \cite{pardo2009pickett}, (BEDT-TTF)$_2$$I_3$ organic salts under pressure \cite{katayama2006suzumura}, photonic metamaterials \cite{wu2014}. However, the only experimental realization for such a dispersion has thus far been observed in optical lattices \cite{leticia2012esslinger}. Because of the possibility to realize the semi-Dirac phases in real materials, it is natural to ask how this anisotropy can be leveraged to understand Hall viscosity in such systems- a question that has received no attention to date. 
\begin{figure}
	\includegraphics[width=0.99\linewidth]{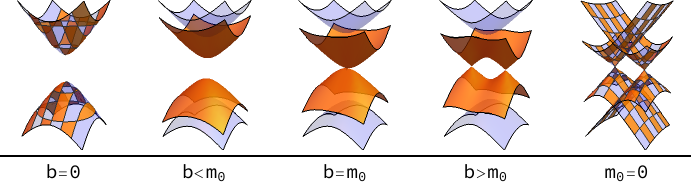}
	\caption{Evolution of energy dispersion of a two-dimensional anisotropic Dirac semi-metal (Eq.~(\ref{eq:DiracHam})) for different values of parameter $b/m_0$. For $b=0$, the spectrum is 
		gapped, whereas for $m_0=0$, we see two gapless Dirac nodes. In both limiting cases the bands are doubly degenerate as presented with cross shaded colors. For $b/m_0=1$, the two Dirac nodes merge, leading to a semi-Dirac point as discussed in the main text.} 
	\label{fig:dispersion}
\end{figure}

The main difficulty to address the above problem comes from the following issue: how does a semi-Dirac material with electrons that have relativistic motion in one Direction and non-relativistic motion along the perpendicular direction couple to the underlying geometry? Given the non-relativistic Hamiltonian (Eq. \eqref{eq:SemiDirac}), there is no straightforward answer to this question. As a result, we propose a different path based on a generalized relativistic model that exhibits three distinct phases, including the critical semi-Dirac phase as a function of an anisotropic parameter. Writing this Hamiltonian on a torus, we find the Landau levels and corresponding wavefunctions. We then derive the formula for Berry curvature with the help of these wavefunctions. We furthermore show how anisotropy leads to {\it more than one independent} Hall viscosity coefficient. Finally, we analyse the scaling of those coefficients as a function of the applied magnetic field and obtain the power law behavior at the critical semi-Dirac phase. These constitute the central results of this paper.
 
{\em Model and phases.--}We begin with the low-energy Hamiltonian of an anisotropic 2D Dirac semi-metal 
\begin{align}
H =  \gamma^0(\mathbf p \cdot\boldsymbol \gamma + {\bf b}\cdot {\boldsymbol\gamma}\gamma^5+m_0).
\label{eq:DiracHam}
\end{align}
Here $\gamma^{\mu}=(\tau^x,-i\tau^y {\boldsymbol \sigma})$, $\gamma^5=\tau^z$ are $4\times4$ Dirac matrices, satisfying $\{\gamma^{\mu},\gamma^{\nu}\}=2\eta^{\mu\nu}\mathbf 1_4$, where $\eta=(1,-1,-1,-1)$; $\tau$ and $\sigma$'s are the Pauli matrices in spin and pseudo-spin space, respectively, 
${\bf p}=(p_x,p_y,0)$; $m_0$ denotes mass gap and  ${\bf b}=(b,0,0)$ is the anisotropic parameter of the Hamiltonian.  
 
The energy spectrum of Eq.~(\ref{eq:DiracHam}) is given by
\begin{align}
E(p_x,p_y)=r\sqrt{p_x^2+p_y^2+b^2+m_0^2+2s\sqrt{b^2(m_0^2+p_x^2)}},
\label{enspec}
\end{align}
where $r,s=\pm$. Note that $r=\pm1, s=-1$ correspond to lowest conduction and highest valence band, respectively.  
The competition between mass gap $m_0$ and anisotropy $b$ leads to three distinct phases (c.f Fig.~\ref{fig:dispersion}). For $b>m_0$, the spectrum is gapless with two-Dirac nodes at $(\pm\sqrt{b^2-m_0^2},0,0)$, while $b<m_0$ corresponds to a gapped insulating phase. On the other hand, for $b=m_0$, we obtain a critical phase where two Dirac nodes merge and lead to a semi-Dirac phase. Thus, the variation of $b/m_0$ changes the Fermi surface topology, leading to a Lifshitz transition.

{\em Landau levels and Wavefunctions on a torus.--}
Let us now focus on finding out the Landau spectrum and corresponding wavefunctions of  Eq.~(\ref{eq:DiracHam}) on a torus. The metric of the torus is given by 
\begin{align}
ds^2=\frac{V}{\tau_2}\left(\mathrm{d}x^2+2\tau_1\mathrm{d}x\mathrm{d}y+ |\tau|^2\mathrm{d}y^2\right),
\end{align}  
where $\tau=\tau_1+i\tau_2$ is the modular parameter and $V$ is the volume of the torus. With this, the Landau Hamiltonian in the presence of a constant perpendicular
magnetic field $B=\epsilon^{ij}\partial_i A_j$  \footnote{We define the epsilon tensor as $\epsilon^{ij}=\frac{1}{\sqrt{g}} \varepsilon_{ij}$ where $\varepsilon_{12}=1$ is the totally antisymmetric symbol.} is obtained to be

\begin{figure}
\includegraphics[width=0.99\linewidth]{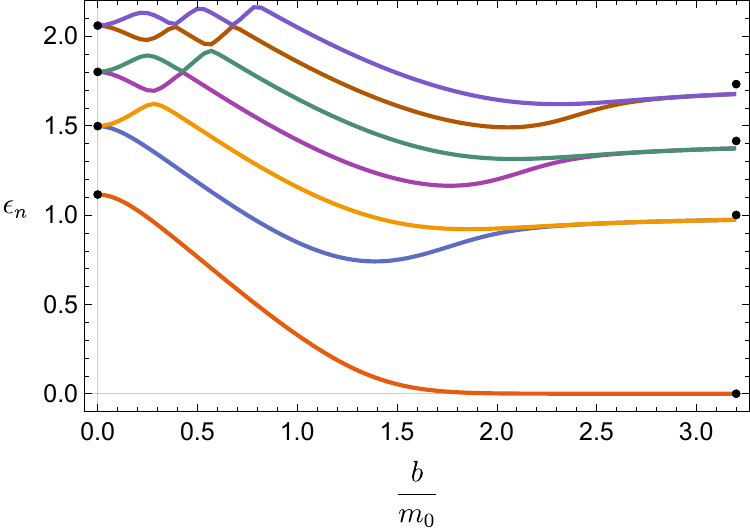}
\caption{Positive energy Landau levels $(n)$. The zeroth Landau level is non-degenerate irrespective 
of the values of $b$ for finite $m_0$. The black dots for $b=0$ correspond to the Landau energies $\epsilon_n\sim\sqrt{n+M^2}$, while black dots for $b/m_0\gg 1$ correspond to $\epsilon_n\sim\sqrt{n}$, showing agreement  between numerical and analytical results.}
\label{fig:LL}
\end{figure}

\begin{align}
H_L = \gamma^0(\Pi_i e_a\,^i\gamma^a+m_0+ b_i e_a\,^i\gamma^a\gamma^5),
\label{eq:LL}
\end{align}
where $a,i,j\in(1,2)$,  $\Pi_i= p_i-e A_i$, and $e_a\,^i$'s are the frame vectors satisfying $g^{ij} = e_a\,^i\delta^{ab} e_b\,^j$. With this construction,
the kinematical momenta $\Pi_{i}$ satisfy $[\Pi_i,\Pi_j]=i\epsilon_{ij}l_B^{-2}$, where $l_B=\hbar/(eB)$ is the magnetic length. 
 
To diagonalize Eq.~(\ref{eq:LL}), we introduce ladder operators $a,a^{\dagger}$, satisfying $[a,a^{\dagger}]=1$. This leads to 
\begin{align}
H_L=\omega\left[(a\sigma_++a^{\dagger}\sigma_-)\tau_z+(\bar d\sigma_++d\sigma_-)\tau_0+M\sigma_x\tau_0\right]\,,
\label{eq:LanHam}
\end{align}
where a bar denotes complex conjugation, $\sigma_{\pm}=\sigma_x\pm\sigma_y$, $\omega=\sqrt{2}l_B^{-1}$, $a=i(V\tau_2)^{-1/2}(\Pi_y-  \tau \Pi_x)/\omega$, $d= -i(V\tau_2)^{-1/2}\tau b/\omega $, and
$M= m_0/\omega$.

For nonzero $b$ and $m_0$, Eq.~(\ref{eq:LanHam}) cannot be exactly diagonalized. Although approximate analytical WKB eigenefunctions can be constructed \cite{halperin2009esaki}, these do not allow for the computation of viscosities for all values of $b$ and $m$. Thus, we choose an algebraic semi-analytic method to diagonalize  Eq.~(\ref{eq:LanHam}) and obtain the Hall viscosities. 
To do so, let us introduce new shifted ladder operators $\mathfrak a_d= a+d$, $\mathfrak a_d^\dagger= a^\dagger+\bar d$. This leads to a set of basis states $|n,\alpha,d,\tau\rangle$, satisfying
\begin{equation}
\mathfrak a_d^\dagger \mathfrak a_d |n,\alpha,d,\tau\rangle = n |n,\alpha,d,\tau\rangle\,.
\end{equation}
The index $\alpha=1,\ldots,N$ labels the magnetic degeneracy, which for notational simplicity we ignore in the rest of the Letter. For a detailed discussion on the existence of these eigenvectors and how to impose the proper boundary conditions on the torus see Ref.~\cite{Levay,Fremling2014}. Having these basis, we proceed to expand each Landau level eigenstate as follows
 \be|\psi\rangle=\sum_n \mathbf c_n|n,d,\tau\rangle,\label{eq:statesExpansion}\ee 
 where $\mathbf c_n$ is a set of four components constant fermion, depending only on the values of $\tau,V$. At this point,  the problem of diagonalizing Eq.~(\ref{eq:LanHam}) is translated into the eigenvalues problem of the infinite matrix
\begin{equation}
H_{nm}\mathbf{c}_m=\epsilon_n\mathbf{c}_n\,,\quad\mathrm{where}\quad H_{nm}=\langle n,d,\tau |H_L|m,d,\tau\rangle\,.
\end{equation}
In general $\mathbf c_n$'s have to be obtained numerically by truncating the series at some large enough values of $n$. However, there are two limiting cases in which the diagonalization  process of Eq.~(\ref{eq:LanHam}) can be done analytically, the first and simplest case corresponds to $m_0=0$ (see Fig. \ref{fig:dispersion}). In this case, the Hamiltonian decouples into two 2-bands subsystem which do not interact with each other.
 The eigenenergies turn out to be $\epsilon_n= \pm\omega\sqrt{n}$ for each subsystem, which in turn leads to double degeneracy (in subsystem subspace). Then the wavefunctions for zeroth Landau level of the two subsystems read off
\begin{eqnarray}\nonumber
\psi_0^1&=&\left(1,0,0,0\right)|0,d,\tau\rangle,\\
\psi_0^2&=&\left(0,0,1,0\right)|0,-d,\tau\rangle.
\label{eq:twoConesWF}
\end{eqnarray} 
For transparency, the higher excited wavefunctions are presented in the supplementary materials. 

In contrast, the second analytically solvable case, $b=0$, is slightly more involved because there is no decoupling. After a careful calculation, the eigenenergies are found to be  $\epsilon_n=\pm\omega\sqrt{n+M^2}$. The zero mode here turns out to be nondegenerate and the states with $n>0$ are doubly degenerate. 
The zeroth Landau level wavefunction is 
\be\psi_{0} =\frac{1}{\sqrt{2}}(1,0, \pm1,0)|0,0,\tau\rangle\,.\ee As before, the degenerate excited wavefunctions can be obtained easily and they are presented in the supplementary material for simplicity.
These limiting behaviors of Eq.~(\ref{eq:LanHam}) are expected to be reflected in the Hall viscosities for both $b\gg m_0$ and  $b\ll m_0$, in which a single viscosity exist and can be computed analytically.

{\em Berry Curvature.--}According to the adiabatic response theorem by Feynman and Hellman, the variation of the Hamiltonian gives rise to two contributions to the leading order
\begin{align}
\left\langle\frac{\partial H}{\partial x_j}\right\rangle=\frac{\partial E}{\partial x_j}-\Omega_{ij}\dot{x}_{j},
\label{eq:adiat}
\end{align}
where $x_j$'s are set of parameters of the Hamiltonian. The first term is a result of the energy change of the ground state deformation. The second term is the adiabatic Berry curvature
\begin{align}
\Omega_{ij}=i\left[\frac{\partial}{\partial x_i } \left<\psi  \left| \frac{\partial \psi}{\partial x_j} \right>-  \frac{\partial }{\partial x_j}\left<   \frac{\partial \psi}{\partial x_i} \right| \psi \right> \right],
\label{eq:berry}
\end{align}
which is nonzero if the phase of the state $\psi$ changes along a closed path in the space of deformations. Plugging the eigenstates Eq. (\ref{eq:statesExpansion}) into Eq. (\ref{eq:berry}),
the total Berry curvature can be readily obtained by   
\begin{align}
\label{eq:BerryOmega}\Omega= i\mathrm d\left(\mathbf{c}^\dagger_{n}\cdot\mathrm d\mathbf{c}_{n}\right) +\mathrm d\left(\mathbf{c}^\dagger_{n}\cdot\mathbf{c}_{m}\right)\wedge\mathcal{A}^{nm}+ \mathbf{c}^\dagger_{n}\cdot\mathbf{c}_{m}\mathcal{F}^{nm} \,.
\end{align}
Here repeated indices denote Einstein's notation and the exterior derivative $\mathrm d$ acts on the space expanded by the parameters $\tau, V$. The detailed derivation is shown in the supplementary material.
The explicit form of $\mathcal{A}$ and $\mathcal F$ evaluated at non-deformed torus $\tau=i$ read
\begin{eqnarray}
\nonumber\mathcal{A}^{mn} &=&-\frac{1}{4}\left(\sqrt{m(m-1)} \delta_{m,n-2}\mathrm d\bar\tau\right.\\
&&+\left.\sqrt{(m+1)(m+2)}\ \delta_{m,n+2}\mathrm d \tau\right),\\
\mathcal{F}^{mn} &=&-\frac{i}{4}\left(m+\frac{1}{2}\right)\delta_{m,n} \mathrm  d\tau\wedge\mathrm d\bar\tau\,.
\end{eqnarray}
For $m_0=0$, the first and second terms in Eq.~(\ref{eq:BerryOmega}) identically vanishes since ${\bf c}_{n}$ are independent of $V,\tau$. The only surviving contribution produces the value for the Berry curvature at $\tau=i$ \footnote{See supplementary material.} 
\begin{equation}
\Omega^{pq} = -\frac{i}{4}\left(n+\frac{1}{2}\delta_{n,0}\right)\mathrm d\tau\wedge\mathrm d\bar\tau \delta^{pq}\,,
\end{equation}
where $p,q=1,2$ label the degenerate subspace associated to each subsystem as pointed out before. Evidently, $\Omega$ is diagonal in the subsystem subspace. Thus, we recover the Berry curvature for isotropic Dirac systems using Eq.~(\ref{eq:BerryOmega})\cite{Kimura:2010yi}\footnote{Notice that, for zeroth Landau level, the Hall viscosity corresponding to this Berry curvature differs from the ones computed in Ref.~ \cite{parrikar2014robert,sharafati2016vignale}.}. 
\begin{figure}
	\includegraphics[width=0.99\linewidth]{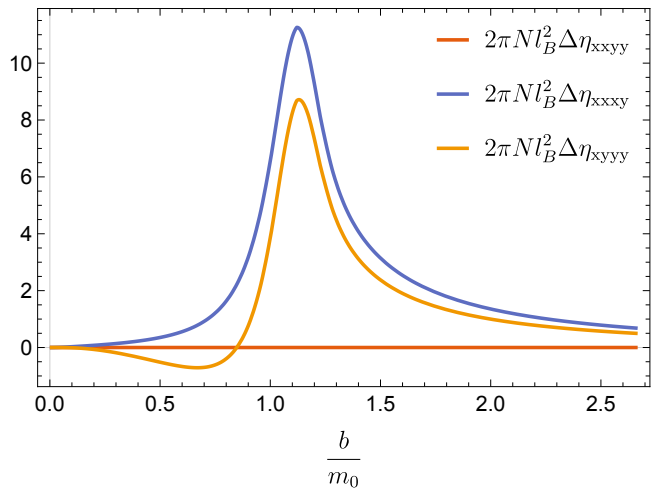}
	\caption{Subtracted viscosity coefficients ($\Delta\eta=\eta-\eta_{\rm{iso}}$) as a function of the ratio $b/m_0$.} 
	\label{fig:Viscosities}
\end{figure}

Similarly, for $b=0$ and for zeroth Landau level ($n=0$), ${\bf c}_n\sim\delta_{n,0}$, leading to  $\Omega=-\frac{i}{8}\frac{\mathrm d\tau\wedge\mathrm d\bar\tau }{\tau_2^2}$. For higher Landau levels ($n>0$), the calculation of Berry curvature is subtle due to two-fold degenerate Landau levels (not to be confused with magnetic degeneracy) as discussed in the preceding sections. 
These subtleties, however, do not change the message we want to convey, because the degeneracy is not present around the critical point. As a result we focus mainly on the zeroth Landau level in our analysis. 
  
For nonzero $b$ and $m_0$, ${\bf c}_n\ne \delta_{n,0}$, thus we may have nonzero contribution from the first and second terms of Eq. \ref{eq:BerryOmega}, 
which in turn may lead to more than one Hall coefficient as will be evident shortly. Thus, this is one of the main results of this study.

{\em Anisotropy and Hall viscosity.--}Armed with the derivation of Berry curvature, we now relate different component of ${\Omega}$ to the viscosity components and show how 
anisotropy in a Dirac system leads to more than one Hall viscosity coefficient. The odd transport coefficients are the most readily visible at the level of constitutive relations. One can expand the average stress tensor in time derivatives of the strain 
\be \label{eq:stress}
T_{ij} = - \sum_{kl}\lambda_{ i j k l}u_{k l}- \sum_{kl}\eta_{ i j k l} \frac{\partial u_{k l}}{\partial t}+\cdots
\ee
where the strain is expressed in terms of a deformation vector $u_{k l}=\partial _k u _l+\partial _l u_k$.
The first term in that expansion $\lambda_{i j k l}$ corresponds to a generalized Hooke's elasticity tensor and the second term $ \eta_{i j k l}$ corresponds to the viscosity tensor. 
Note that $\eta_{ijkl}$ is symmetric under exchange of $i$ with $j$ and $k$ with $l$ \footnote{In a non rotational invariant system the stress tensor is not necessarily symmetric, however in this work we shall consider  only response to the symmetrized strain.}. In general, $\eta$ can be divided into
$\eta=\eta_S+\eta_A$, where $\eta_S$ is symmetric with respect to interchanging first pair $(ij)$ with $(kl)$ whereas $\eta_A$ is antisymmetric under exchange of $(ij)$ with $(kl)$. Since antisymmetric part is odd under time-reversal $\eta_A\ne 0$ only when time reversal symmetry is broken. As the antisymmetric part of the viscosity tensor is non-dissipative may survive at zero temperature. From now on we only focus on the antisymmetric non-dissipative part and  remove the label $A$ for brevity and clarity. 
\begin{figure}
	\includegraphics[width=0.99\linewidth]{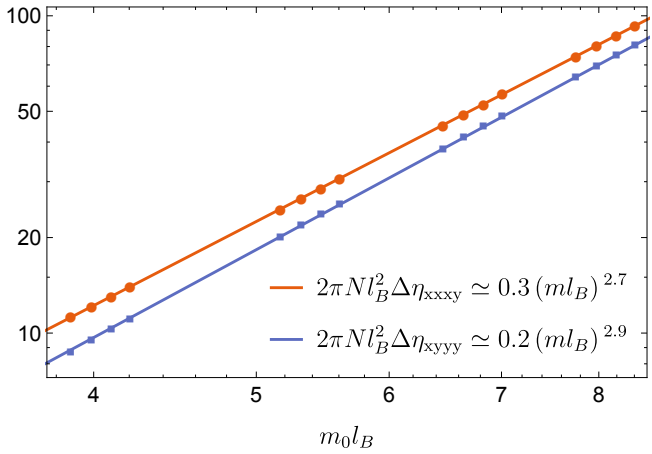}
	\caption{Scaling of the subtracted viscosity tensor as a function of the magnetic length. Dots correspond to the numerical data whereas solid lines represent the fitting.} 
	\label{fig:powerlaw}
\end{figure}

For generic two dimensional (2D) systems without time reversal, $\eta_{ijkl}$ has in principle three independent components $\eta_{xxxy}$, $\eta_{xyyy}$ and $\eta_{xxyy}$.
However, for rotationally invariant systems, the number of independent quantities reduces, and $\eta_{ijkl}$ is solely determined by a single viscosity, denoted as $\eta_H$ since $\eta_H=\eta_{xxxy}=\eta_{xyxx}$ and $\eta_{xxyy}=0$.  
This single object turns out to be a universal quantity $\eta_H=\bar s\rho/2 $, where $\bar s$ is the orbital angular momentum, and $\rho$ is the average number density. 

Since the odd viscosity tensor is a multicomponent tensor for a generic 2D system, we relate different components of Berry curvature to the viscosity coefficients. Together with Eq.~(\ref{eq:adiat}), (\ref{eq:stress}), and evaluating strain rate on the non-deformed torus $\tau=i,V=1$, we obtain 
\bea
2\pi N l_B^2\eta_{xxxy} &=& \Omega_{\tau_1\tau_2}-\Omega_{\tau_1V} \\
2\pi N l_B^2\eta_{xyyy} &=& \Omega_{\tau_1\tau_2}+\Omega_{\tau_1V}\\
2\pi N l_B^2\eta_{yyxx} &=& \Omega_{\tau_2V},
\eea
where each component of $\Omega$ can be extracted from Eq.~(\ref{eq:BerryOmega}) (see supplementary material). For isotropic case, $\Omega_{\tau_2V}$ and $\Omega_{\tau_1V}$ turns out to be identically zero. Thus  $\eta_{xxxy}=-\eta_{yyxy}$ is the only parameter that determines response to the geometry of the QH states. However, due to anisotropic nature of the Dirac system in Eq.~(\ref{eq:DiracHam}), each term of $\Omega$ contribute except $\Omega_{\tau_2V}$ which remains zero (see Fig.~\ref{fig:Viscosities}).

Fig.~(\ref{fig:Viscosities}) illustrates the different components of the viscosity tensor after subtracting the isotropic value ($\Delta\eta=\eta-\eta_{\mathrm{iso}}$)  as a function of the ratio $b/m_0$. 
It is evident that the different components of $\eta$ start to deviate from the universal isotropic value as we increase $b$ for fixed $m_0$ and become maximum near the ideal semi-Dirac phase ($b/m_0=1$). Thus, the anisotropy $b\ne0$ leads to more than single viscosity coefficients, in contrast to the isotropic case ($b=0$). If we further increase $b/m_0$, both nonzero components of $\eta$ start to reduce and merge again to the isotropic value. This is attributed to the fact that in the large $b\gg m_0$ limit, we obtain two  well-separated Dirac nodes, in conjunction with the earlier discussion. Consequently, the wave function behaves approximately as Eqs. \ref{eq:twoConesWF}, which in turn gives the isotropic value of Hall viscosity. 

We next aim to find the dependence of viscosity coefficients on the magnetic field $B$ near the semi-Dirac phase. 
It is known for typical isotropic 2D system, $\eta\sim B$ \cite{avron1995zograf}, irrespective of the relativistic or non-relativistic nature of the electrons. In contrast, we find a different scaling behavior of the subtracted viscosity near the semi-Dirac point. Fig.~\ref{fig:powerlaw} illustrates  the maximum of $\Delta \eta$ for several values of $m\, l_B$. In the given range of analysed data, we can fit the power-law scaling as seen in Fig. \ref{fig:powerlaw}. We observe $\Delta \eta$ goes to zero for large magnetic fields.
This confirms an intuitive picture that for large enough energy scale, the system behaves as isotropic.

 {\em Conclusion.--} We have introduced a framework for studying the non-dissipative transport in anisotropic Dirac semi-metals, where the anisotropy is present due to a preferred direction. This distinguishes this model from the previous cases studied in the literature, where isotropy is broken by a tensor \cite{Haldane2015,gromov2017bradlyn}. We have introduced a relativistic model with an anisotropic vector that reproduces the spectrum of the non-relativistic semi-Dirac system Eq. (\ref{eq:SemiDirac}) at low energies for certain values of parameters. We have derived an universal formula for a Berry curvature in this model that succinctly captures the anisotropy for semi-Dirac. Using the formula, we have numerically investigated how the anisotropy leads to the departure from one Hall viscosity coefficient for the zeroth Landau level. We have shown that at the critical semi-Dirac point, the odd stress tensor has two non-equal entries. In addition to that, we have shown that the third entry is identically zero. 

Our theory allows one to have detailed further studies of Hall transport in anisotropic semi-Dirac systems. The theory is covariant and can be used to perform systematic studies based on effective actions and geometric responses. 

Finally the studies presented here can be useful to generalise the existing isotropic, parity-odd hydrodynamic solutions \cite{Lucas:2014sia,Scaffidi2017,Delacretaz:2017yia,Ganeshan2017} to the anisotropic systems. This would supplement existing analysis that takes into account dissipative viscosities \cite{Link:2017ora}.

{\em Acknowledgments.--}F.P-B would like to express a special thanks to the Mainz Institute for Theoretical Physics (MITP) for its hospitality and support. We would like to thank Andrey Gromov, Karl Landsteiner, Mar\'ia Vozmediano, Alberto Cortijo, Piotr Witkowski and especially Barry Bradlyn for enlightening discussions. This work was supported by the Deutsche Forschungsgemeinschaft via the Leibniz Programm.

\bibliographystyle{apsrev4-1}
\bibliography{AnisotropicHall}
\onecolumngrid
\section{Supplementary Material}	
\subsection{Anisotropic Berry curvature}
In general the computation of Berry curvature demands the knowledge of the Hamiltonian's eigenstates. However, diagonalising a system is not always under full analytical control. In this section, we will use the algebraic properties of the Ladder operator eigenstates to write a general formula, which happens to be useful for the numerical computation of  Berry curvature. As a first step, we introduce the set of Ladder operators
\begin{eqnarray}
\mathfrak a_b &=& i(2V\tau_2)^{-1/2}l_B(\Pi_y+b_y-  \tau (\Pi_x + b_x))\\
\mathfrak a_b^\dagger &=& -i(2V\tau_2)^{-1/2}l_B(\Pi_y+b_y-  \bar\tau (\Pi_x + b_x)),
\end{eqnarray}
satisfying the commutation relations $[\mathfrak a_b,\mathfrak a_b^\dagger ]=1$, where $[\Pi_x,\Pi_y]=iVB$. On the torus, the magnetic flux is quantized $l_B^{-2}V=2\pi N$. Given these operators, there is a basis obeying
\begin{equation}
\mathfrak a_d^\dagger \mathfrak a_d |n,\alpha,d,\tau\rangle = n |n,\alpha,d,\tau\rangle\,, \qquad t_1^N|n,\alpha,d,\tau\rangle=|n,\alpha,d,\tau\rangle\,,\qquad t_2^N|n,\alpha,d,\tau\rangle=|n,\alpha,d,\tau\rangle\,,
\end{equation}
where $t_{1,2}^N$ are magnetic translations along the two cycles of the torus (For details on this topic and the specific form of these operators, see \cite{fremling,Levay}). 
The vacuum states are defined  as follow
\begin{eqnarray}
\mathfrak a_d|0,\alpha,d,\tau\rangle = 0 \,, \qquad t_1^\beta|0,\alpha,d,\tau\rangle=e^{-2\pi i\frac{\beta n}{N}}|0,\alpha,d,\tau\rangle\,,\qquad t_2^\beta|0,\alpha,d,\tau\rangle=|0,\alpha-\beta,d,\tau\rangle\,.
\end{eqnarray}
The Ladder operators and the magnetic translations commute, which allow us to construct the whole tower of degenerate states
\begin{eqnarray}
|n,\alpha,d,\tau\rangle = \frac{1}{\sqrt{n!}}(\mathfrak a_d^\dagger)^n|0,\alpha,d,\tau\rangle \qquad,\qquad \langle m,\beta,d,\tau|n,\alpha,d,\tau\rangle = \delta_{m,n}\delta_{\beta\alpha}\,,
\end{eqnarray}
where the meaning of the label $\alpha$ is clarified now. It corresponds to the magnetic degeneracy and takes values  $\alpha=1,2,\ldots,N$. 

At this point, we can use this basis to expand the anisotropic four component fermion for a given Landau level as
\begin{equation}
|\psi_\alpha\rangle = \sum_n  \mathbf{c_n}|n,\alpha,d,\tau\rangle
\end{equation}
with $\mathbf{c_{n}} = \left(c^1_{n},c^2_{n},c^3_{n},c^4_{n}\right)$. For simplicity, we assume the Landau levels do not have other degeneracy than the magnetic one. Assuming that such state exist its Berry connection read
\begin{equation}
A_{\alpha\beta} = i\langle\psi_\alpha|{\mathrm d}|\psi_{\beta}\rangle = i\delta_{\alpha\beta}\sum_{n}\left(\mathbf{c^*_{n}}\cdot\mathrm d\mathbf{c_{n}}\right) + \sum_{n,m}\left(\mathbf{c^*_{n}}\cdot\mathbf{c_{m}}\right)\mathcal{A}_{\alpha\beta}^{nm}
\end{equation}
where we have introduce the following one form
\begin{equation}
\mathcal{A}_{\alpha\beta}^{nm} = i\langle n,\alpha,d,\tau|{\mathrm d}|m,\beta,d,\tau\rangle \,,
\end{equation}
which corresponds to the Berry connection of Schr\"odinger particles when $n=m$ \cite{levay}.  To compute the one form $\mathcal{A}_{\alpha\beta}$, we will follow an algebraic approach based on \cite{levay,hoker,vinet}. In doing so, we introduce the generators of $SL(2,\mathcal{R})$ 
\begin{eqnarray}
J_1 &=& \frac{1}{8\pi N}\left((p_y+b_y)^2-(p_x+b_x)^2\right)= -\left.\frac{1}{4}\left(\mathfrak a_d\mathfrak a_d+\mathfrak a_d^\dagger\mathfrak  a_d^\dagger\right)\right|_{\tau=i},\\
 J_2 &=& -\frac{1}{8\pi N}\left((p_x+b_x)(p_y+b_y)-(p_y+b_y)(p_x+b_x)\right)=-\left.\frac{1}{4i}\left(\mathfrak a_d\mathfrak a_d-\mathfrak a_d^\dagger\mathfrak a_d^\dagger\right)\right|_{\tau=i},\\
  J_3 &=& -\frac{1}{8\pi N}\left((p_y+b_y)^2+(p_x+b_x)^2\right)=-\left.\frac{1}{2}\left(\mathfrak a_d^\dagger \mathfrak a_d+\frac{1}{2}\right)\right|_{\tau=i}\,,
\end{eqnarray}
is not difficult to check that they satisfy the commutation relations
\begin{equation}
[J_1,J_2] = -iJ_3\qquad,\qquad [J_2,J_3] = i J_1 \qquad,\qquad [J_3,J_1] = i J_2.
\end{equation}
All these properties allow us to write the following relation
\begin{equation}
\frac{1}{2}\left(\mathfrak a_d^\dagger\mathfrak a_d+\frac{1}{2}\right)=X_1J_1-X_2J_2-X_3J_3= \mathcal U(-J_3)\mathcal U^{-1}
\end{equation}
with $\mathcal U = \exp\left(-i\rho\left(\sin\phi J_1-\cos\phi J_2\right)\right)$ and the following sequence of change of coordinates \cite{note}
\begin{eqnarray}
X_1 &=& \frac{1-|\tau|^2}{2\tau_2} \qquad,\qquad X_2 = \frac{\tau_1}{\tau_2} \qquad,\qquad X_3 = \frac{1+|\tau|^2}{2\tau_2} \,,\\
X_1 &=& \sinh\rho\cos\phi  \qquad,\qquad X_2 = \sinh\rho\sin\phi \qquad,\qquad X_3 = \cosh\rho\,.
\end{eqnarray}
Therefore, we conclude that the harmonic oscillator eigenstate are unitarily related to the eigenstates of $J_3$, which we call $|n,\alpha,d,i\rangle$ 
\begin{equation}
|n,\alpha,\tau\rangle =\mathcal U|n,\alpha,i\rangle.
\end{equation}
Finally the unitary operator $\mathcal U$ allows us to write $\mathcal A_{\alpha\beta}$ as follows
\begin{equation}
\mathcal{A}^{nm}_{\alpha\beta}  =i \langle n,\alpha,d,i| \mathcal U^{-1}{\mathrm d}\mathcal U|m,\alpha,d,i\rangle\,,
\end{equation}
Given the fact that $\mathcal U^{-1}{\mathrm d}\mathcal U$ is an element of the Lie algebra $SL(2,\mathcal R)$, it can be expanded in terms of $J_1,J_2,J_3$ and the coefficients are one forms. These one form are independent of the representation used for the generators. Therefore, we use Pauli matrices, $(J_1,J_2,J_3)=(\sigma_1/2i,\sigma_2/2i,\sigma_3/2)$ to compute them
\begin{eqnarray}
\omega_1 &=& \sin\phi\mathrm{d}\rho+\cos\phi\sinh\rho\mathrm{d}\phi\,,\\
\omega_2 &=& -\cos\phi\mathrm{d}\rho+\sin\phi\sinh\rho\mathrm{d}\phi\,,\\
\omega_3 &=& -\left(1-\cosh\frac{\rho}{2}\right)\mathrm{d}\phi\,,
\end{eqnarray}
\begin{equation}
\mathcal{A}^{nm}_{\alpha\beta}  =\left(\omega_1\langle n,\alpha,d,i| J_1|m,\alpha,d,i\rangle+ \omega_2\langle n,\alpha,d,i |J_2| m,\alpha,d,i\rangle + \omega_3\langle n,\alpha,d,i |J_3| m,\alpha,d,i\rangle\right)\,,
\end{equation}

where the expectation values can be computed explicitly using the ladder operator's properties
\begin{eqnarray}
\langle n,\alpha,d,i| J_1|m,\alpha,d,i\rangle &=& -\frac{1}{4}\left(\sqrt{m(m-1)}\delta_{n,m-2}+\sqrt{(m+1)(m+2)}\delta_{n,m+2}\right)\\
\langle n,\alpha,d,i |J_2| m,\alpha,d,i\rangle &=& -\frac{1}{4i}\left(\sqrt{m(m-1)}\delta_{n,m-2}-\sqrt{(m+1)(m+2)}\delta_{n,m+2}\right)\\
\langle n,\alpha,d,i |J_3| m,\alpha,d,i\rangle&=& -\frac{1}{2}\left(m+\frac{1}{2}\right)\delta_{n,m}\,.
\end{eqnarray}
Finally $\mathcal{A}^{nm}_{\alpha\beta}$ in terms of the coordinates $\tau,\bar\tau$ reads
\begin{eqnarray}\nonumber
\mathcal{A}^{nm}_{\alpha\beta} &=& -\frac{1}{4\tau_2}\left[ \frac{ 1-i \tau }{ 1+i\bar\tau}\mathrm{d}\bar \tau\sqrt{m(m-1)}\delta_{n,m-2}+\frac{1+i\bar\tau }{1-i\tau }\mathrm{d} \tau\sqrt{(m+1)(m+2)}\delta_{n,m+2} +
\left(\mathrm d\bar\tau+\mathrm d\tau + \mathrm{d}\Lambda\right)\left(m+\frac{1}{2}\right)\delta_{n,m}\right]\delta_{\alpha\beta},\\
\end{eqnarray}
where $\Lambda=-4 i \tanh ^{-1}\left(\frac{\tau_1}{1+\tau_2}\right)$. 
Taking the exterior derivative of the Berry connection, we obtain the Berry curvature
\begin{equation}
\label{eq_BerryOmega}\Omega_{\alpha\beta} = \mathrm{d}A_{\alpha\beta} = i\delta_{\alpha\beta}\sum_{n}\mathrm d\left(\mathbf{c^*_{n}}\cdot\mathrm d\mathbf{c_{n}}\right) + \sum_{n, m}\mathrm d\left(\mathbf{c^*_{n}}\cdot\mathbf{c_{m}}\right)\wedge\mathcal{A}_{\alpha\beta}^{nm}+ \sum_{n, m}\mathbf{c^*_{n}}\cdot\mathbf{c_{m}}\mathcal{F}_{\alpha\beta}^{nm}, 
\end{equation}
where again we have introduce the two form $\mathcal{F}_{\alpha\beta}^{nm}=\mathrm{d}\mathcal{A}_{\alpha\beta}^{nm}$ 
\begin{eqnarray}
\mathcal{F}^{nm}_{\alpha\beta}  &=& \frac{\mathrm{d}\tau\wedge\mathrm{d}\bar \tau}{8i\tau_2^2}\left[ \frac{ 1-i \bar\tau }{1+i\bar\tau}\sqrt{m(m-1)}\delta_{n,m-2}+\frac{1+i\tau }{1-i\tau } \tau\sqrt{(m+1)(m+2)}\delta_{n,m+2} +
\left(m+\frac{1}{2}\right)\delta_{n,m}\right]\delta_{\alpha\beta}.
\end{eqnarray}
Thus we recover Eqs. (14)-(16) of the main text.

\subsection{Hall viscosities and Berry curvature}
The aim of this section is to derive Eqs. (19)-(21) of the main text. For this derivation we follow \cite{avron}. Applying strain on a physical system, is equivalent  at the linear order to deforming the background space
\begin{equation}
g_{ij} = \delta_{ij}+u_{ij},
\end{equation}
where $\delta_{ij}$ is the non-deformed metric and $u_{ij}$ the strain tensor.
This interpretation  allows us to relate the strain rates with the parameters $\tau,V$ as follows
\begin{equation}
\mathrm d u_{11} = \frac{1}{2}\left(\frac{\mathrm dV}{V}-\frac{\mathrm d\tau_2}{\tau_2}\right) \quad,\quad \mathrm du_{22} = \frac{1}{2}\left(\frac{\mathrm dV}{V}+\frac{\mathrm d\tau_2}{\tau_2}\right) \quad,\quad \mathrm du_{12} = \frac{1}{2}\frac{\mathrm d\tau_1}{\tau_2}\,.
\end{equation}
On the other hand, the Berry curvature in general will have the following form
\begin{equation}
\Omega = \Omega_{\tau_1\tau_2} \mathrm d\tau_1\wedge\mathrm d\tau_2 +  \Omega_{\tau_1V} \mathrm d\tau_1\wedge\mathrm dV + \Omega_{\tau_2V} \mathrm d\tau_2\wedge\mathrm dV \,,
\end{equation}
which can be written in terms of the  physical variables (strain) as
\begin{equation}
\Omega = 2\tau_2(\tau_2\Omega_{\tau_1\tau_2}-V\Omega_{\tau_1V})\mathrm du_{11} \wedge \mathrm du_{12} + 2\tau_2(\tau_2\Omega_{\tau_1\tau_2}+V\Omega_{\tau_1V}) \mathrm du_{12}\wedge\mathrm du_{22} - 2\tau_2 V \Omega_{\tau_2V}  \mathrm du_{11}\wedge\mathrm du_{22}.
\end{equation}
Finally from these relations, we extract the components of the Hall viscosity
\begin{eqnarray}
\eta_{xxxy} &=& \frac{ B}{2\pi N}(\tau_2^2\Omega_{\tau_1\tau_2})-\tau_2\Omega_{\tau_1V} \\
\eta_{xyyy} &=& \frac{ B}{2\pi N}(\tau_2^2\Omega_{\tau_1\tau_2})+\tau_2\Omega_{\tau_1V}\\
\eta_{xxyy} &=& -\tau_2\Omega_{\tau_2V} 
\end{eqnarray}

\subsection{Massless phase}
As discussed in the main text, the Landau Hamiltonian under study can be exactly diagonalized when $m_0=0$. In this case the Hamiltonian reads
\begin{align}
H_L=\omega\left[(a\sigma_++a^{\dagger}\sigma_-)\tau_z+(\bar d\sigma_++d\sigma_-)\tau_0\right]\,.
\label{eq:LanHamsuppl0}
\end{align}
The system at this specific value for the mass ($m_0$)  simplifies due to the fact that it decomposes into two decoupled systems because the Hamiltonian is block diagonal. This fact implies that all eigenenergies are doubly degenerate with values
\begin{equation}
\epsilon_n = \pm\omega\sqrt{n}\,,\qquad n\ge 0.
\end{equation}
The eigenstates are
\begin{eqnarray}\nonumber
\psi_n^1&=&\frac{1}{\sqrt{2}}\left[(\sqrt{2}-1)\delta_{n,0}+1\right]\left(|n,\alpha,d,\tau\rangle,\pm |n-1,\alpha,d,\tau\rangle,0,0\right)\\
\psi_n^2&=&\frac{1}{\sqrt{2}}\left[(\sqrt{2}-1)\delta_{n,0}+1\right]\left(0,0,|n,\alpha,-d,\tau\rangle,\pm |n-1,\alpha,-d,\tau\rangle\right).
\end{eqnarray}
In practice, the decoupling implies that a Berry curvature can be associated to each subsystem, in other words the Berry curvature is diagonal in the degeneracy subspace
\begin{equation}
\Omega^{pq}_{\alpha\beta} = -\frac{i}{4}\left(n+\frac{1}{2}\delta_{n,0}\right)\frac{\mathrm d\tau\wedge\mathrm d\bar\tau }{\tau_2^2}\delta^{pq}\delta_{\alpha\beta}\,,
\end{equation}
where $p,q=1,2$ label the band degeneracy and $\alpha,\beta=1,2,\ldots,N$ the magnetic degeneracy.


\subsection{Purely insulating phase}
The second case, in which the Hamiltonian can be exactly diagonalized,  corresponds to $b=0$. At this special value of the parameters, the Landau Hamiltonian reads
\begin{align}
\label{eq:LanHamsuppl1}
H_L=\omega\left[(a\sigma_++a^{\dagger}\sigma_-)\tau_z+M\sigma_x\tau_0\right]\,.
\end{align}
The Eq.~(60) leads to a nondegenrate eigenstate for $n=0$ and doubly degenerate states for $n>0$. This is in contrast to the degeneracy discussed in the preceding section as Eq.~(60) cannot be decomposed into two decoupled subsystems. This fact have important implications on the Berry curvature. In fact, the Berry curvature turns  non-abelian and not diagonal in degeneracy subspace. In particular, the eigenergies are
\begin{equation}
\epsilon_n = \pm\omega\sqrt{M^2+n}\,,\qquad n\ge 0.
\end{equation}
The non-degenerate zeroth Landau level is
\begin{eqnarray}
|\psi_{0} \rangle &=& \frac{1}{\sqrt{2}}(1,0, \pm1,0)|0,\alpha,\tau\rangle,
\end{eqnarray}
and the degenerate states are
\begin{eqnarray}
|\psi_{n,1} \rangle &=&\sqrt{\frac{1}{2}}\frac{1}{\lambda^2+n}(\left(n-\lambda^2\right)|n,\alpha,0,\tau\rangle,0, \left(\lambda^2+n\right)|n,\alpha,0,\tau\rangle,-2 \sqrt{n} \lambda |n-1,\alpha,0,\tau\rangle),\\
|\psi_{n,2} \rangle&=&\sqrt{\frac{1}{2}}\frac{1}{\lambda^2+n}(2 \sqrt{n} \lambda|n,\alpha,0,\tau\rangle,(\lambda^2+n)|n-1,\alpha,0,\tau\rangle,0,(n-\lambda^2)|n-1,\alpha,0,\tau\rangle)\,,
\end{eqnarray}
where $\lambda=\pm\sqrt{n+M^2}-M$. 

Due to subtleties in computing Berry curvature for the degenerate case $n>0$, we only present it for zeroth Landau level, which turns out to be 

\begin{equation}
\Omega_{\alpha\beta} = -\frac{i}{8}\frac{\mathrm d\tau\wedge\mathrm d\bar\tau }{\tau_2^2}\delta_{\alpha\beta}\,,
\end{equation}
where again $\alpha,\beta=1,2,\ldots,N$ label the magnetic degeneracy.

	

\end{document}